# The Virtual Experiences Lab
# – a platform for global collaborative engineering and beyond

**Ian D. Peake, Jan Olaf Blech, Ian Thomas, Nicholas May,
Heinz W. Schmidt, Lasith Fernando, Ravi Sreenivasamurthy**

RMIT University, Melbourne, Australia
{ ian.peake | janolaf.blech | ian.edward.thomas |nicholas.may
| heinrich.schmidt | lasith.fernando | ravi.sreenivasamurthy } @rmit.edu.au

**INTRODUCTION**

We are developing the Virtual Experiences (Vx)Lab, a research and research training infrastructure and capability platform for global collaboration. VxLab comprises labs with visualisation capabilities, including underpinning networking to global points of presence, videoconferencing and high-performance computation, simulation and rendering, and sensors and actuators such as robotic instruments locally and in connected remote labs. VxLab has been used for industry projects in industrial automation, experimental research in cloud deployment, workshops and remote capability demonstrations, teaching advanced-level courses in games development, and student software engineering projects. Our goal is for resources to become a "catalyst" for IT-driven research results both within the university and with external industry partners. Use cases include: multi-disciplinary collaboration, prototyping and troubleshooting requiring multiple viewpoints and architectures, dashboards and decision support for global remote planning and operations, e.g. in automation, power, facilities or environmental management, and safe student access to remote or dangerous labs without "suiting up". VxLab was originally developed (as "VITELab") to support research collaboration within the Australia-India Research Centre for Automation Software Engineering (AICAUSE) [2,3,4], a partnership between ABB and RMIT University. ABB is a global leader in power and automation systems. In this talk we describe VxLab and its main use cases.

The *Global Operations Visualization (GOV) Lab* supports local user collaboration while accessing and displaying standard applications and internet services through a high resolution 8m x 2m video display wall. Local services include video conference (H323 / skype) and streaming to/from local and remote sites. GOV Lab is designed for use with parallel rendering middleware such as SAGE middleware [1]. The figure below shows a user at the GOV Lab visualization wall.

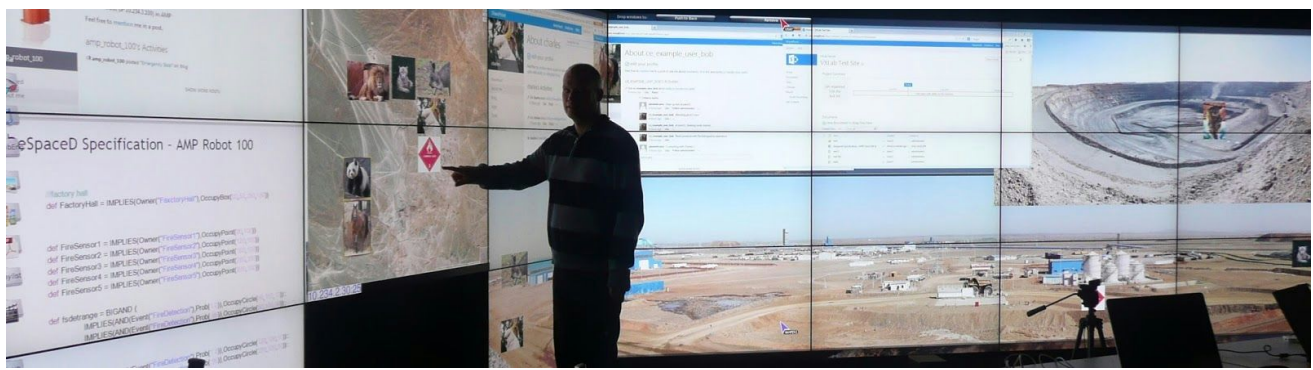

The *Cyber-Physical Simulation (CS)Rack* provides an experimental 640 core HP blade server, to support modeling, simulation and collaboration services. CSRack is publicly accessible and configured via OpenStack. Our intention is to explore the use of CSRack as a "cloudlet," locally-connected, low-latency and high-bandwidth, for use as a gateway to national e-research infrastructure such as NeCTAR. The *Advanced Manufacturing Robotic Interoperation Test (AMRIT) Lab* provides two ABB IRB120 industrial robot arms with three-fingered grippers, sensors and cameras. The VxLab Network is logically separate from the RMIT corporate network to create a sandbox environment. Dedicated private network links have been built to RMIT labs and to ABB's Melbourne and Bangalore sites (separate from the ABB corporate network).

**CHIMINEY**

The Chiminey system [5] is a cloud-based computing platform that enables scientists to perform complex computation on cloud-based and traditional high performance computing (HPC) facilities. Chiminey gives special importance to resource access and management abstraction, allowing seamless execution on NCI, MASSIVE, NeCTAR, Amazon AWS and other platforms. Scientists are not expected to have a technical understanding of cloud computing, HPC, or fault tolerance in order to leverage the benefits provided by Chiminey. MyTardis compatibility is included for automatic rganisation, curation and analysis and plotting of results. Chiminey has been ported experimentally to CSRack with a view to providing a simplified front end for CSRack HPC users.

**WS-REDUNDANCY**

CSRack is being used to execute a number of controlled tests for redundancy experiments. These experiments test a redundancy protocol (WS-Redundancy) that allows greater control over the non-functional attributes of groups of redundant services: an early version of the protocol is described by May et al. [11]. CSRack provides an ideal environment in which to control the deployment, connection, and execution of test runs across a range of distributed services. In addition, it supports a flexible test infrastructure, embodied as redundancy, testing, and data management servers.

**COLLABORATIVE ENGINEERING**

Collaborative Engineering [8,9] is a platform developed in collaboration with ABB that aims at facilitating operation and maintenance of remote mining and industrial plant operations in Australia. Its core comprises a service bus reacting to events / alarms generated by remote facilities. Displaying appropriate information to stakeholders in reaction to events is realized using VxLab. In particular, we are using VxLab's visualization capabilities and the computing infrastructure for offering web services for communication of external devices such as robots and machinery as well as deciding on possible reaction and selecting appropriate data for human evaluation using our BeSpaceD spatial decision framework [10].

**OTHER USAGE, CUSTOMIZATION AND EXTENSIONS**

Usage has steadily increased since 2013 with student projects starting December 2014, lab demonstrations and demonstrations by others using the lab [7]. We use virtual desktops to integrate applications into the video wall including high resolution web browsers. Multiple users can (e.g.) collaboratively display slide presentations, view live video from robot lab, hold video conferences, and work with IDEs such as ABB's Robot Studio to develop and modify robot code. With NTNU Trondheim we have ongoing research work in remote use of lab facilities in areas such as model-driven software development for distributed embedded systems. With researchers in media and communications we are developing the *VxPortal cluster*, enabling immersive, reconfigurable multi-perspective stereo 3D views of virtual exhibits and scenes (a successor of VROOM [6].) Finally researchers in property construction and project management are developing a decision support system on a curved video projection wall in VxLab.

**ABOUT THE AUTHOR(S)**
Dr Ian Peake is a Research Fellow at RMIT in the School of Computer Science and Information Technology since 2007 and manager of the VXLab since 2012. His research interests are in software engineering, in particular in component-based software architecture of real time and parallel systems. Since 1992 he has worked in industry-focused research with companies such as ICL/Fujitsu, Oracle and ABB at the University of Queensland, Monash University and RMIT University. Dr Peake gained his PhD in 2002.

Dr Jan Blech is a research fellow at RMIT. He has (co-)authored more than 40 peer-reviewed publications in the areas of software engineering, formal methods and embedded systems. Dr. Blech has worked on several publicly funded projects and collaborated with major automotive and industrial automation companies in Europe. His experience comprises research and technical project leadership roles.

Ian Thomas is a software developer and system administrator at the eResearch Office of RMIT University. He has worked in data curation for output of high-performance computing systems, microscopy data for materials, and screen media objects (film and television). His current work is in institutional metadata stores, decision support systems for climate change modelling, and in cloud-based platforms in support of eResearch applications.

Nicholas May is a software engineer with the eResearch Office at RMIT University. He is a certified professional member of the Australian Computer Society, and has more than twenty five years of software engineering experience across the software development lifecycle. In addition, he is a PhD Candidate in the School of Computer Science & Information Technology at RMIT University, with research interests in the fields of service-oriented computing and fault tolerance.

Heinz Schmidt is Professor of Software Engineering at RMIT University, in Melbourne Australia, where he heads Distributed Software Engineering and Architecture in the School of Computer Science and directs the e-Research Office. He received his PhD from Bremen University, Germany. He is also Adjunct Professor at Maelardalen University, in Vaesteras, Sweden, and at Monash University, Melbourne, Australia. Prof Schmidt is internationally recognised in software engineering for parallel and distributed systems, notably systems composed of large numbers of interacting components in application areas ranging from industrial automation to telecommunication networking and artifial intelligence. He has over 30 years experience with object-oriented and component-based software architectures, languages, systems and tools in this domain, in practice, research and training. Heinz has published over 120 refereed articles on aspects of distributed and concurrent systems using Petri Nets, automata, logic, abstract data types, graphs and other methods, and on practical methods and tools for constructing and testing such systems. He has published several books and book chapters.

Lasith Fernando is a software engineer within AICAUSE. He is an embedded systems software engineer by profession and his areas of expertise include Smart Energy Management Systems in residential and commercial context, active RFID technology sensor networks and contactless secure payment terminals. He is currently driving his career towards the automation and control systems engineering arena with a special interest in industrial automation and robotics. From 2010-2012 he worked as an embedded software engineer at the Centre for Technology Infusion at La Trobe University. Lasith obtained his BSc in Computer Science from University of Colombo, Sri Lanka in 2006 and MSc in Applied Electronics from the same university in 2010.